# Surface Vacancy Generation by STM Tunneling Electrons in the Presence of Indigo Molecules on Cu(111)

Carlos Javier Villagómez,* Fernando Buendía, Lauro Oliver Paz-Borbón, Bernardo Fuentes, Tomaso Zambelli, and Xavier Bouju

**Abstract**

Herein, we investigate the consequence of local voltage pulses on the adsorption state of single indigo molecules on the Cu(111) surface as well as on the atomic structure underneath the molecule. With a scanning tunneling microscope, at 5 K, intact molecules are imaged as two lobes corresponding to the electron density of each indoxyl moiety of the molecule which are connected by a carbon double bond. Then, two short successive voltage pulses with the tip placed above the molecule generate irreversible modifications, as revealed by consecutive scanning tunneling microscopy (STM) imaging. Density-functional theory calculations coupled to STM image calculations indicate the creation of a double surface vacancy of copper surface atoms below the oxygen atom of the indigo molecule as the most plausible scenario. These extracted copper atoms are stabilized as adatoms by the indigo oxygens, oxidizing each copper adatom to 0.32 electron.

**Introduction**

The indigo molecule is an ancient organic dye molecule that has been used for centuries to dye colorful blue color textiles, and they even appear in recipes of Babylonian transcripts.[1] Nowadays, it is widely used as a pigment to dye billions of blue jeans per year.[2] On the other hand, due to the electronic, optical, and chemical properties of the indigo molecule and its derivatives, several applications are explored such as field-effect transistors,[3,4] solar cells,[5,6] diodes,[7-9] memory storage devices,[10] diesel markers,[11] DNA biosensors,[12] or chemical detectors of $NO_2$.[13] Also, the chemical structure of indigo motivates exploring new chromophores with greater optoelectronic features.[14,15] There exist wide literature reports on indigo and its derivatives with regard to the high photostability of the compounds. Three phenomena are invoked to account for an intramolecular change causing photostability: (1) photoexcitation of an excited state of the molecule for trans–cis isomerization through the C═C bond; (2) isomeric change from the keto to monoenol configuration (Figure 1) by excited-state intramolecular proton (ESIP) transfer, where the hydrogen atom of the nitrogen atom is transferred to the oxygen; finally, (3) the isomeric dienol configuration resulting from excited-state dual proton (ESDP) transfer, where both hydrogen atoms of the amine central groups are transferred to both oxygen atoms of the carbonyl groups. The first mechanism appears less favorable due to steric crowding between hydrogen atoms in the final form and the important interaction of the hydrogen bonding between the carbonyl and N–H groups.[16-20] However, some thioindigo derivatives show trans–cis isomerization and the involvement of a triple state in the photoisomerization[21,22] and particularly N,N′-di(t-butoxycarbonyl)indigo[23] shows the absence of intramolecular N–H⋯O bonding. The photostability of indigo is thus been attributed to the ESIP transfer from the keto to enol intramolecular isomerization after irradiation (Figure 1), where the proton transfer occurs from the nitrogen to the oxygen atoms at the femtosecond scale, and an internal conversion allows the recovery of the keto conformation at the picosecond scale.[24-26] Briefly, after light irradiation: (a) indigo is excited to the electronic state $S_1$ ($1\pi\pi^*$) with a different potential energy surface than the $S_0$ state; (b) after relaxation of the molecule, the intramolecular proton transfer occurs at the femtosecond scale; (c) then, there is a fast non-radiative decay by motion of the molecule in the $S_1$ minimum toward the conical intersection of the $S_1/S_0$ states, and internal conversion is reached; and (d) the molecule returns to the $S_0$ ground state minimum from the enol to keto form. The fast non-radiative decay of indigo via hopping is the most accepted mechanism of photostability. This process explains why indigo is robustly photostable in the gas phase.[16,19,24-35]

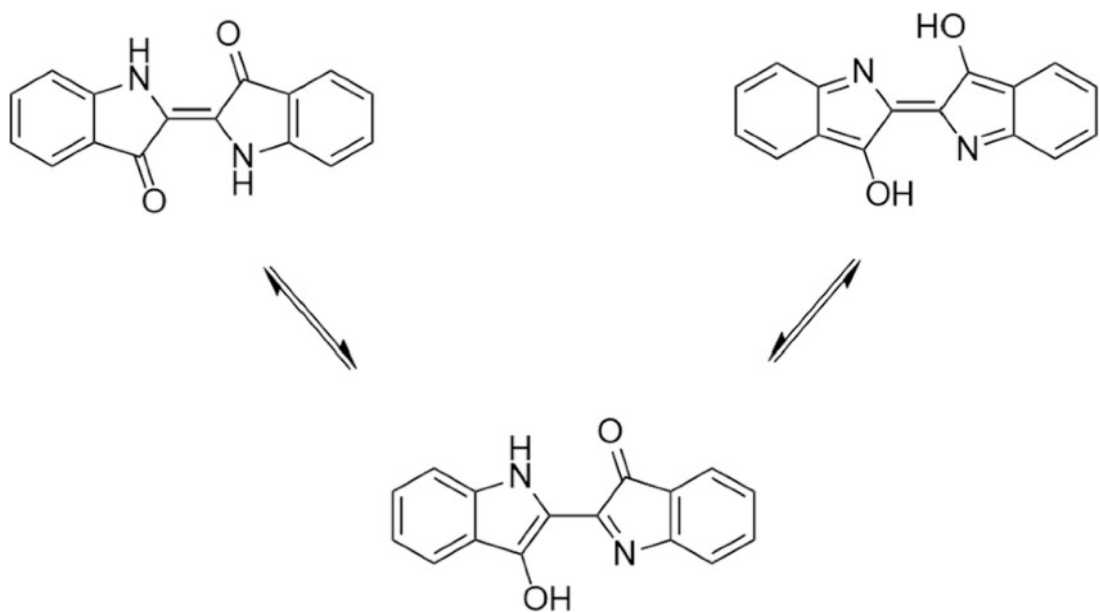

*Figure 1. Left, representation of the diketo indigo molecule (simply named indigo in the following). Bottom, representation of the monoenol indigo after a single proton transfer. Right, representation of the dienol indigo molecule after a double proton transfer mechanism. Notice that the double central C–C bond for indigo changes to a single bond for the monoenol indigo, whereas it remains double for the dienol indigo.*

Because the photostability is triggered by the proton transfer in the gas phase and in solution, one can ask what happens when molecules are adsorbed on a substrate. Indeed, since a long time ago, the photostability of indigo has been assigned to the moiety of the molecule with the hydrogen bonding between the carbonyl and N–H groups, and the presence of a surface may affect this mechanism. Nevertheless, most of the studies on adsorption of indigo on novel materials for applications in semiconductor organic electronic devices have focused on the ambipolar charge transfer mobility and the frontier molecular orbital alignment (HOMO–LUMO) with respect to the Fermi level on metal electrodes. Theoretical calculations for understanding the chemistry of the molecule also show that the reactivity of indigo is highly affected by substituents of the N–H group.[36,37] However, there are few studies on indigo adsorbed onto surfaces tackling the optical and electronic properties of the dye. STM experiments carried out by Honda et al.[38] studied the alkyl derivatives of the indigo copper complex deposited on the o-dichlorobenzene/graphite interface; they found out that the molecules exhibit a mordanting effect on the surface (addition of metal atoms to stabilize the adsorption of dye molecules) in comparison to the molecules without the metal ions and observed the formation of 2D and 3D chiral structures for larger alkyl chains. In our previous work,[39] we studied the indigo molecules by STM at different deposition rates on

Cu(111). We have reported that molecules assemble in linear chains at low coverages presenting chirality due to the symmetry reduction on the surface. Indigo forms almost homochiral chains by hydrogen bonding. At higher coverages, the molecules presenting the other chirality compared to the ones in the almost homochiral chains are expelled, and indigo molecules form enantiopure domains.

The stabilization of the mono- and dienol forms seems to be challenging in the gas phase and solution, but it remains achievable for molecules adsorbed on a surface. On the other hand, electron excitation carried out by the STM tip pulses is expected to induce isomerization or intramolecular proton transfer (single or dual).[40,41] In comparison to azobenzene where the trans–cis isomerization can be carried out by optical excitation or even induced by electric-field excitation of STM on molecules adsorbed on the surface, indigo does not show such an isomerization of the C═C bond by photoexcitation.[17,24]

The motivation of the present study was not to study the optical properties of the molecule adsorbed on the metallic surface but to address the possible ESIP transfer as a response of the indigo molecule to a local electron stimulus instead of photoexcitation. Once adsorbed onto a surface, a localized electron injection could generate the permanent monoenol state from indigo by ESIP transfer due to the molecule/surface interaction. Then, a second stimulus would lead to the dienol form. An STM tip can be used for such a manipulation of individual adsorbates. It is now well established that STM manipulations can be performed according to three different strategies. First, by using mechanical lateral forces between the tip and the adsorbate, one can pull, slide, or push a molecule or an atom between definite positions on the surface.[42-52] Second, the localized electric field, enhanced by the tip effect, can induce a controlled motion of the adsorbate,[40,53-55] as exemplified by the recent nanocar race.[56-61] Finally, the current of energetic tunnel electrons can chemically modify the conformation of a molecular adsorbate.[41,62-66] Thus, by triggering the excitation strength, a local-probe tip is able to prompt conformational changes[67,68] and even breaking/formation of chemical bonds.[49,69-77]

Applying localized voltage pulses with an STM tip on the keto group of molecules on Cu(111) at 5 K, we reproducibly observed the irreversible transformation of the molecules. We considered two possible mechanisms to explain these facts: (i) a stable keto–enol transfer stabilized by the surface and (ii) the formation of a vacancy on the copper surface with consequent formation of a complex between the oxygen atoms of the molecule and the metallic adatoms extracted from the surface.

For instance, previous manipulations of molecules induced by electrons from STM pulses showed tautomerization and proton transfer of a free base tetraphenyl-porphyrin (TPP) on Ag(111) between two states for 2H-TPP or among four states after removing one internal hydrogen atom of the TPP molecular core.[41] Molecular orientation induced by tunneling electrons has been shown to switch azobenzene molecules adsorbed on the Au(111) surface at the threshold voltage applied to the molecules.[78] Then, a similar question arises for the indigo molecule about the possible reaction induced by the excitation of the molecule adsorbed on the metal surface by the tunneling electrons of the STM tip. In this way, it is improbable that indigo adsorbed on the surface carries out isomerization by the central C═C bond, but uncertainty exists as to whether the pulses can induce a permanent isomeric tautomerization of the molecule. In this work, we concentrate on the adsorption at the single-molecule level to understand the individual adsorption of the molecule and the manipulation by the STM tip pulses.

**Experimental and Computational Methods**

The Cu(111) crystal surface (Goodfellow Inc., UK) was cleaned by standard repeated procedures of Ar+ sputtering, followed by annealing of the sample to 600 °C at a UHV pressure of $10^{-10}$ Torr. The indigo molecules were deposited by sublimation using an external filament with molecules adsorbed onto it, the evaporator was positioned pointing the sample placed inside a low-temperature cryostat by an access hole, while the surface crystal was kept at a low temperature in order to have single molecules during the adsorption process and to avoid diffusion of the molecules for autoassembling. All the measurements were carried out at a low temperature of 5 K and at an UHV chamber pressure of $10^{-11}$ Torr.

The tungsten STM tips were fabricated by electrochemical attack of a 0.25 mm diameter wire immersed in a solution of NaOH, and after the fabrication, the STM tips were annealed by Joule heating inside the preparation chamber to desorb the possible oxidation product formation in air. The cleanliness of the surface and the STM tip was verified by STM image measurements in situ showing larger flat terraces and monoatomic step edges.

Density functional calculations were performed using Vienna ab initio simulation package 5.4 (VASP 5.4).[79-81] The gradient-corrected Perdew–Burke–Ernzerhof (PBE) exchange and correlation functional[82] was used in combination with the Tkatchenko–Scheffler van der Waals (vdW) dispersion.[83] Projected-augmented wave potentials were used for different species (Cu, C, N, O, and H atoms). The Brillouin zone was sampled in a $k$-mesh of $3 \times 3 \times 1$ employing an energy cutoff of 400 eV for the plane waves. The copper substrate was a Cu(111) fcc $3 \times 3$ slab with three atomic layers in the non-periodic direction. The lattice parameter employed was 3.63 Å, obtained after slab relaxation without restrictions and this value is similar to those inferred in previous experimental studies.[84] On the other hand, for the adsorption process of the indigo molecule, the deepest atomic layer was kept fixed to mimic the behavior of the inner atomic layers of the solids. The convergence criterion for the electronic self-consistent field was $1 \times 10^{-4}$ eV, whereas the energetic threshold was $1 \times 10^{-3}$ eV for the relaxation process. Finally, STM images were calculated with the extended Hückel molecular orbital-elastic scattering quantum chemistry (EHMO-ESQC) code.[85,86] Here, the STM junction comprising the substrate, the adsorbates, the tip apex, and the tip support is fully described at the atomic level. A set of semiempirical extended Hückel orbitals is assigned on each atom site. The scattering of tunnel electrons through the junction is calculated to allow the evaluation of the tunneling current with the Landauer formula. This method has already proven its reliability with small and large molecular systems.[85,87-90]

**Experimental Results**

On STM images, an isolated molecule consistently appears like a dumbbell with two nearly circular lobes (Figure 2a).[39] This defines the main axis of the molecule relying on the direction of the two rings. The bright spots have a relative height of about 0.7 Å and a distance separation of 1.1 nm. As previously mentioned, STM pulses were used in experiments in the past to induce different conformations of individual molecules on the surface due to tautomerization or mechanical movement on the terraces and step edges. After imaging the intact indigo molecule, a voltage pulse of 2.4 V was applied on one lobe of the molecule (blue spot in Figure 2a) for a time of 400 ms. We ensured that the regulating loop was switched off before the pulse and restored after the pulse time. To obtain quantitative information, we applied systematic other pulses at various bias voltages on indigo molecules to induce conformational changes. We have chosen the tip–surface distance corresponding to the setpoint (10 pA, −300 mV) as the initial point $Z_o$, and we applied pulses with varying durations until the changes were observed. In Figure 3a, experimental data are shown corresponding to the duration threshold to successfully change the appearance of the molecule for each bias voltage.

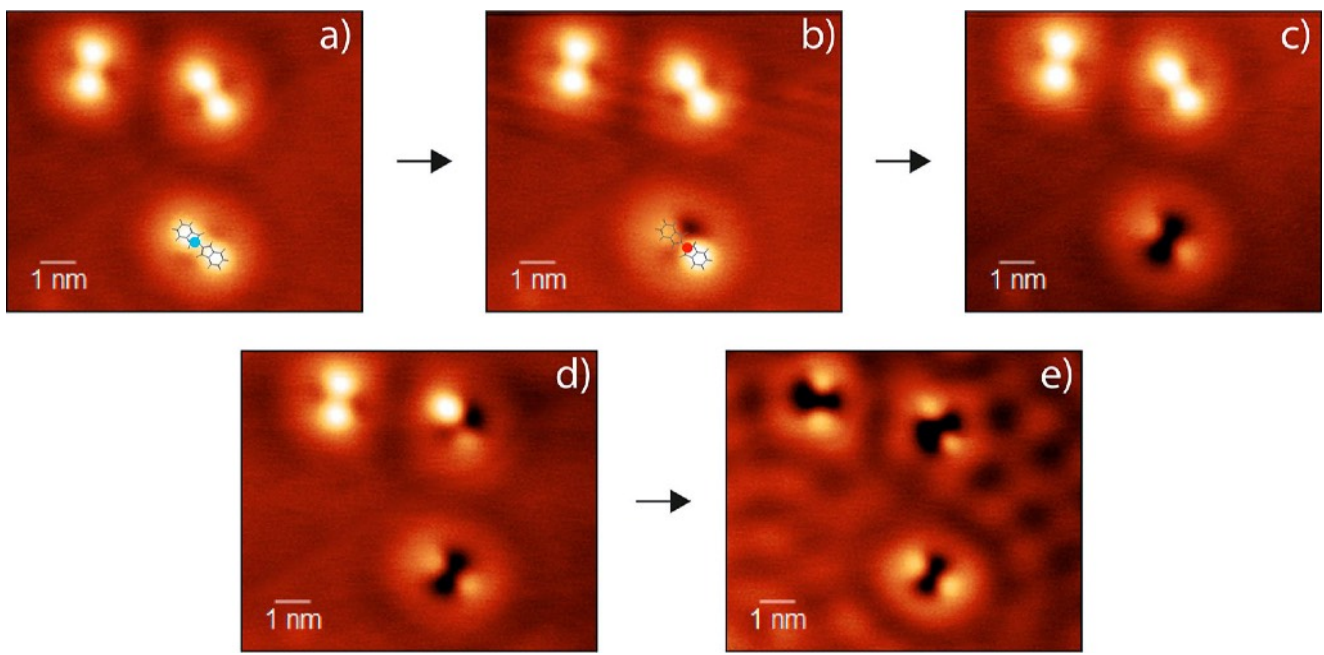

*Figure 2.* (a) Three indigo molecules on Cu(111) prior to any voltage pulse; (b) one molecule is turned into a new form after a first positive voltage pulse with the tip positioned between the two lobes, marked by the blue spot in (a); (c) a second pulse is applied, marked by the red spot in (b) changing the conformation of the targeted molecule; (d,e) the same manipulation process is done on the other two molecules (STM imaging: 10 pA, −300 mV).

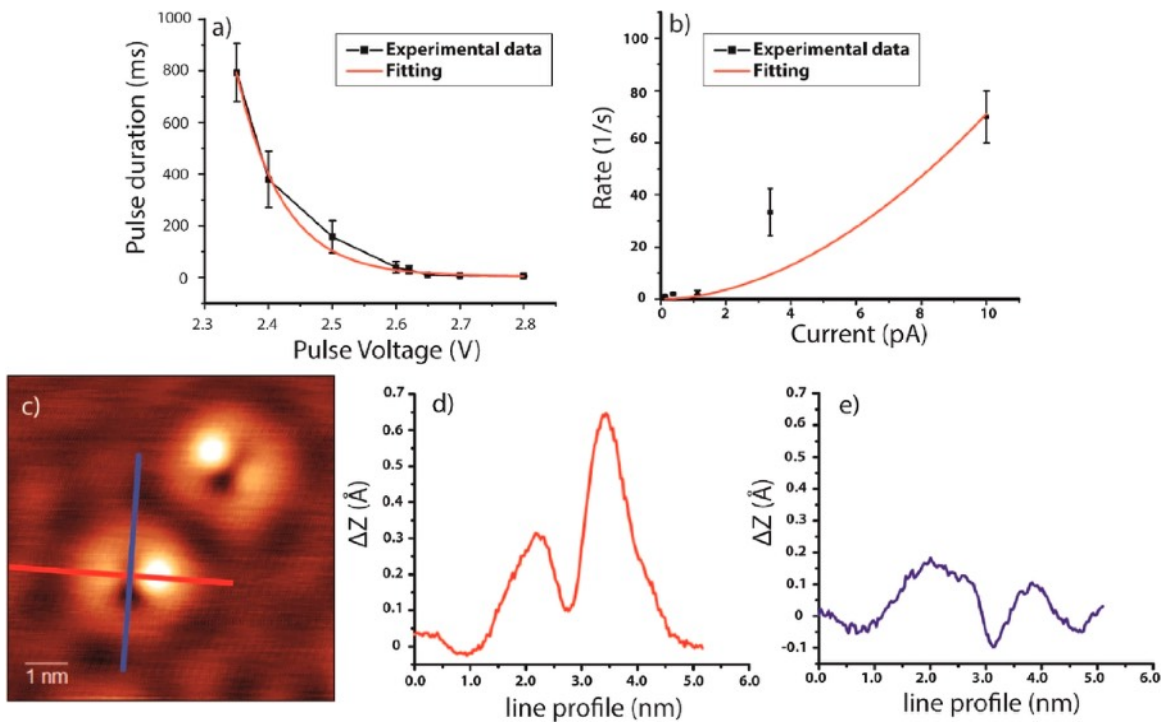

*Figure 3*. (a) Quantitative information about the switching mechanism: duration threshold from which a conformational change is obtained as a function of the pulse voltage. The fitting red curve has the form $A \cdot exp(-eV/B)$ with $A = 1.26 \times 10^{15}$ ms and $B = 83$ meV. (b) Manipulation rate as a function of tunneling current at 2.4 V. (c) STM images of two individual molecules after a single manipulation. (d,e) Scan lines along the red and blue lines in (b), respectively.

The curve follows an exponential decrease that can be fitted by $A \cdot \exp(-eV/B)$ with $A = 1.26 \times 10^{15}$ ms and $B = 83$ meV. From that, one can deduce that the dose, that is to say, the number of tunneling electrons impinging on the molecule and the surface, varies as $N = N_0 \cdot \exp[-e(V - V_0)/B]$, with $N_0 = I_0 A/e$ ($I_0 = 10$ pA) and $V_0 = -300$ mV. The dose varies exponentially, meaning that the more the voltage bias, the more energetic the electrons, and the more the efficiency change. Additionally, the manipulation rate $R$ versus the tunneling current $I$ follows a power-law dependence $R \propto I^\alpha$ using a model for bond breaking by inelastic electron tunneling,[62,91,92] which is reported in Figure 3b for a bias voltage of 2.4 V (for each tip–sample distance, one deduces a set point current and measures the transformation rate). One can deduce that the parameter $\alpha = 1.85 \pm 0.05$, indicating that one deals with a double-electron process.[93,94] The STM image exhibits a contrast change as shown in Figure 2b (molecule at the bottom) corresponding to a conformational change of the molecule. After this first bias voltage pulse was applied above a targeted molecule, the dumbbell feature disappears (Figure 3c): a bright lobe is switched off (relative height diminished to 0.3 Å, Figure 3c), and a lateral depression becomes much darker (a contrast of about 0.2 Å with respect to the central part of the molecule is measured as one can see with the blue profile in Figure 3d). Then, after a second pulse of 3 V (red spot in Figure 2b), the remaining bright spot is turned off to 0.3 Å, and a pronounced dark trace appears perpendicularly to the main axis (Figure 2c) with a depression around 0.2 Å relative to the central molecule contrast similar to the one shown in Figure 3d. Notice that this double process does not apparently modify the location of the molecule like a rotation or a translation onto the surface, as shown in Figure 2d,e after several STM manipulations proceeded on two other molecules. We observed that the second voltage pulse has to be higher than the first one, meaning that the second manipulation requires a higher electron energy. Moreover, only positive bias voltage pulses are efficient to induce a manipulation, a fact that we are not able to rationalize.

**Analysis and Discussion**

The goal of this work is thus to determine which phenomenon is responsible for the observed change of contrast of the indigo molecule upon STM pulses. For this purpose, calculations are a necessary complement to the experiments, and the first step is the determination of the most stable adsorption site of the indigo molecule on the Cu(111) substrate by means of DFT calculations. Once this has been determined, we continue to calculate the energy barrier of the hydrogen transfer and the conformational structure of the monoenol and dienol chemical structures on the surface. Then, calculated STM images with the electronic scattering quantum chemistry (ESQC) method are obtained for each structure for comparison with experimental images.

The initial DFT calculations show that the indigo molecule is adsorbed parallel to the surface of the Cu(111) slab. The preferred orientation of the molecule was analyzed after the relaxation of the molecule adsorbed on the slab in 12 different orientations, each of them rotated 15° with respect to the previous one ($n \times 180°/12$, with $n = 0, ..., 11$), as shown in Figure 4. The rotation axis is located above a bridge position of the Cu(111) surface. This position was found to be more energetically favored at the top or hollow positions. The $C_{2h}$ symmetry of the molecule and the location of the rotation axis enabled the angles greater than 165° to generate equivalent geometric configurations. The inclusion of the dispersion energy is crucial for these systems due to the interaction between the aromatic rings of the molecule and the metal substrate, and it was addressed using the Tkatchenko–Scheffler vdW dispersion correction.[83] We have measured various orientations of adsorbed molecules and have checked that they are compatible with the metastable and stable orientations in Figure 4. Once the preferred orientation of the indigo ground state molecule (GS) is determined, the possible reactions generated by the voltage pulse were explored. The proton transfer from some other possible reactions, such as the double-hydrogen transfer or the generation of Cu vacancies was taken into account to rationalize the phenomenon observed in the experimental STM images. The single proton transfer barrier was obtained by means of the nudge elastic band (NEB) method, and five images between the initial and final states were employed. The final state was determined in an independent search, and it is corroborated as a local minimum in the potential energy surface. Bader analysis was performed for different systems using the method developed by the Henkelman group[95] to get a deeper insight of the electronic structure of the molecule–substrate complex in various conformations.

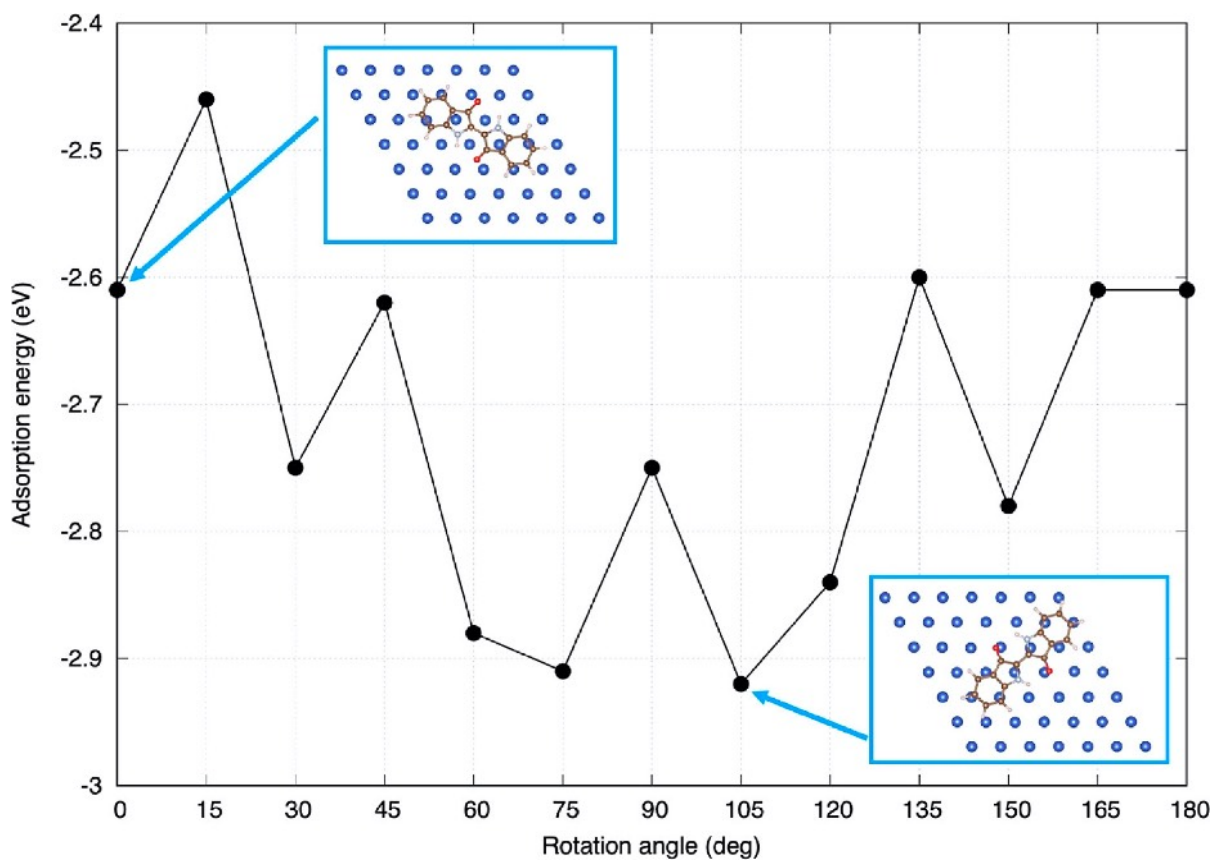

*Figure 4. Adsorption energy of the indigo molecule with respect to its orientation above the Cu(111) surface.*

The adsorption energy of the molecule on the Cu(111) substrate was obtained with the following equation: $E_{ads} = E_{indigo/Cu(111)} - E_{indigo} - E_{Cu(111)}$, where $E_{indigo/Cu(111)}$ represents the total energy of the relaxed geometry of the adsorbed indigo on the Cu(111) substrate, while $E_{indigo}$ and $E_{Cu(111)}$ are the total energies of the relaxed molecule and the substrate, respectively.

The activation barrier of the protonation process of the molecule on Cu(111) is calculated as the difference $E_{indigo/Cu(111)\text{-TS}} - E_{indigo/Cu(111)\text{-init}}$, where $E_{indigo/Cu(111)\text{-TS}}$ represents the total energy of the adsorbed indigo geometry in the transition state (TS) of proton transfer.

Finally, the energy to generate a Cu vacancy was obtained by means of the following equation: $E_{indigo/Cu(111)} - E_{indigo/Cu(111)+vac} - E_{Cu}$, where $E_{indigo/Cu(111)+vac}$ represents the total energy of the geometry of the indigo supported on the copper substrate with the Cu vacancy, while $E_{Cu}$ is the total energy of the copper atom alone.

As mentioned already, due to the $C_{2h}$ symmetry and a similar interaction of the indoxyl groups with the Cu(111) surface slab, the following explanations are limited for half of the molecule and can be applied for both sides. Previously, semiempirical calculations were performed without relaxation of the surface atoms and have provided similar qualitative results compared to the present DFT ones.[39] However, for accurate determination of transient states, DFT calculations were preferred in the following. Here, DFT calculations presented in Figure 4 show that the central C=C bond sits above a bridge site of the (111) surface. The most stable conformation (GS) is found at 105° with respect to the reference, and it possesses an adsorption energy of −2.92 eV, its geometric structure is at the bottom of Figure 4. The less stable system (15°) has an adsorption energy of −2.46, and it is 0.46 eV higher than that of the most stable molecule–substrate conformation. Another adsorption conformation with a high stability is the molecule orientated at 75° from the reference, and it is only 0.01 eV less stable than GS. For all cases, the molecule strongly interacts with the surface by means of the oxygen atom, and this bond generates the distortion of the molecule and the substrate. For the most stable geometry, the interaction of the oxygen atom with the substrate generates a decrease in the C–O interaction compared with the gas-phase molecule, which is reflected in the elongation of the C–O bond length by 0.09 Å. The C–O chemical bond decreases from a double to a single bond; for heterocyclic aromatic compounds with a carbonyl group, it is achievable to bond metals through a lone pair of electrons.[96] With regard to the Cu atom beneath the oxygen atom, it experiences a slight lift of 0.14 Å with respect to its original height. This phenomenon occurs on the two indoxyl groups of the molecule. In accordance with our results, other DFT calculations of indigo molecules adsorbed on silver nanoparticles show the formation of a complex between the molecule and the substrate via the O–Ag interaction for different silver clusters of $Ag_2$, $Ag_{14}$, and $Ag_{16}$.[97-99] In addition to the flat-lying adsorption preference of the molecule on the silver substrate,[97] it is also worth noticing that surface-enhanced Raman spectroscopy measurements of indigo molecules adsorbed on silver nanoparticles irradiated at 514 nm involve intramolecular electronic excitations, intercluster excitations, and charge-transfer effects. These effects are visualized in the intensity enhancement and appearance of new vibrational Raman spectra, but it is not the case for other laser excitation wavelengths of lower energy where just the charge-transfer effect is involved.[99] This energy value is very close to our electronic tunneling excitation of 2.4 eV pulses with the copper surface.

On the other hand, the vdW interaction between the aromatic rings of the molecule and the substrate is essential for the adsorption process.[100] The total energy values including a vdW interaction range from −2.46 to −2.92 eV, while without the vdW interaction they vary from −0.31 to 0.64 eV. This difference reflects the importance of taking into account the dispersion interactions for complexes formed by organic molecules and metallic substrates. For the most stable system, the distances between the center of the rings and the surface are 2.54 and 2.56 Å. On the other hand, the N–H bond length (1.01 Å) shows a similar value compared to the gas phase (1.02 Å). The hydrogen bond between C–O and N–H is similar to the gas phase, but the distance between the two groups is slightly increased by 0.04 Å due to the introduction of the substrate. The charge analysis corroborates the ionic character of the O–Cu bond with a charge transfer of 0.19 electron (e-) from the Cu toward the oxygen. However, the charge in the oxygen atom is similar to that in the gas phase due to the charge received from the C atom in the gas phase. As a result of the new Cu–O interaction, the C=O double bond is weakened to a single bond, this is confirmed by the change in the bond length. These results show that the vdW dispersion and C–O interactions are responsible for the geometric configuration of the molecule–Cu(111) substrate complex.

The finding of the most stable position and orientation of the molecule on the Cu(111) substrate is crucial as a starting point in the search for the possible reactions that could occur in the system after the STM pulse. Various mechanisms can be explored to explain the resulting molecular contrast change in the STM images after the pulse, such as keto–enol tautomerization, trans–cis isomerization, fragmentation, dissociation, and surface vacancy generation. In the following, only an intact molecule without bond breaking is considered because no fragments have been observed. Moreover, a partial molecular dissociation consisting of the detachment of the oxygen atom that could be adsorbed nearby the molecule on a hollow site has been estimated and appeared largely unfavorable compared to the mechanisms discussed after.

The first explored mechanism is the possibility of a permanent proton transfer process on the surface due to the electronic and chemical interactions of the indigo molecule with the substrate in contrast to solution irradiation experiments. Such a keto–enol tautomerization mechanism has already been proposed for the Au/isophorone complex, where the surface and the presence of gold adatoms play a crucial role in the appearance of an enol tautomer.[101] The trans–cis isomerization was not explored because this process requires higher energy due to the strong interaction of the indoxyl group with the copper surface. The switching of other aromatic double bond compounds lying flat on the surface such as azobenzene on Au(111) was possible by adding chemical spacers such as *tert*-butyl in order to separate the phenyl rings from the surface interaction.[40] This is not the case for the indigo compound that is directly adsorbed on the surface.

The chemical process of one hydrogen transfer from nitrogen to oxygen is presented in Figure 5, showing that the symmetry constraint of the indigo molecule is broken once the hydrogen is transferred. The reaction was analyzed using the most stable configuration such as the initial state N–H, shown in Figure 5, and the molecule after the hydrogen transfer that has been obtained in an independent search such as the final state O–H, shown in Figure 5. On the other hand, the TS (in Figure 5) was found using the NEB method. In the gas phase, once the proton has been transferred, the C=O double bond goes from 1.24 to 1.28 Å, showing the weakening of the bond due to the new O–H bond. On the other hand, the distance between N and C atoms close to the center of the molecule is shortened, from 1.37 Å in the initial state to 1.30 Å once the hydrogen has been transferred. The reaction on the Cu(111) substrate presents some differences with respect to the gas phase case. The Cu–O bond is broken due to the new O–H interaction, and it is reflected in the Cu–O bond length that increases from 2.12 to 2.64 Å after the hydrogen transfer. Moreover, the nitrogen atom does not generate a new bond with a Cu surface atom, in contrast with the oxygen atom before the hydrogen transfer. The rings slightly change their geometry, but the distances between the aromatic rings and the substrate are similar. On the other hand, the Cu atom of the substrate that was bound with the oxygen atom before the proton transfer returns to its original position. The new system is 0.28 eV higher in energy than the initial state. Bader analysis exhibits a charge transfer of 0.14 e⁻ from the substrate toward the nitrogen of the molecule after the hydrogen transfer. This value is smaller than the charge received by the oxygen before the hydrogen transfer, and this difference is related to the higher electronegativity of the oxygen atom. The double hydrogen-transfer system was also analyzed, and the geometric changes are similar to both sides of the molecule. This system shows lower stability by 0.79 eV compared with the most stable system, and it is 0.51 eV higher in energy than the single hydrogen transfer. The C–O bond distance on both sides of the molecule (1.30 Å) is increased compared with the one hydrogen-transfer system (1.28 Å) and also with respect to the gas phase (1.24 Å). On the other hand, the two Cu atoms close to the nitrogen atoms do not change their original position, while the two Cu atoms close to the oxygen atoms return to the original height of the remaining Cu atoms in the slab.

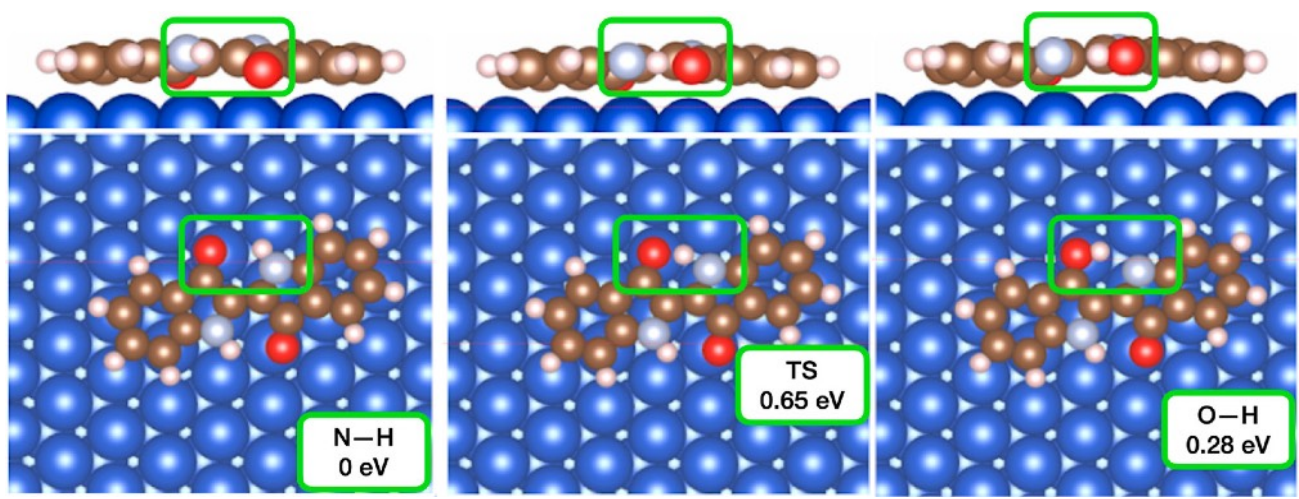

***Figure 5.*** *Top and side views of the molecule conformation during the proton transfer from the nitrogen atom to the oxygen atom. The energy barrier to cross the transient state (TS) is 0.65 eV with a final state at + 0.28 eV compared to the initial state.*

The minimum energy path (MEP) of the hydrogen transfer was analyzed by using eight intermediate geometries between the initial and the final state. The TS is 0.65 eV higher in energy than the initial state as can be seen in Figure 5. This value is lower than the one observed for the system in the gas phase. The geometry of the TS locates the hydrogen atom at 2.94 Å from the closest Cu atom of the substrate, 1.35 Å from the nitrogen and 1.40 Å from the oxygen atom. The oxygen atom is located at 2.49 Å from the closer Cu atom of the surface, while the nitrogen one is 2.90 Å from the closer Cu atom of the surface. The copper atom close to the oxygen atom presents a lift from the surface of 0.10 Å compared with their neighbors. This fact reflects the contribution of the substrate to stabilize the intermediate state due to a Cu–O interaction, while the transferred hydrogen atom seems to be interacting very weakly with the substrate.

The new metastable configurations present lower stability compared with the indigo molecule adsorbed in the ground state; however, they could be accessible configurations after the pulse introduction. In order to verify if the proposed systems are observed experimentally after the pulse, ESQC calculations of the calculated DFT geometries were carried out: (1) keto, (2) monoenol, and (3) dienol, in Figure 6a–c, respectively. The STM calculated images in Figure 6 show a surprisingly negligible change of the STM contrast. By comparing Figure 6b,a, for a single proton transfer, almost insignificant modification is observed, and only a tiny indentation near a lobe is visualized. For the double proton transfer, there is slight diminution in the intensity surrounding the center of the lobes, Figure 6c. However, the experimental STM images (Figure 2b,c) differ greatly with respect to the calculated STM images with one or two proton transfers: there are no dark depressions in the middle of the molecule. The hydrogen-transfer process has been analyzed in depth as a possible path in view of the fact of previously reported studies and also because the STM voltage pulses applied on the molecule on the substrate could help induce the reaction. Nevertheless, the single and double proton transfers are discarded.

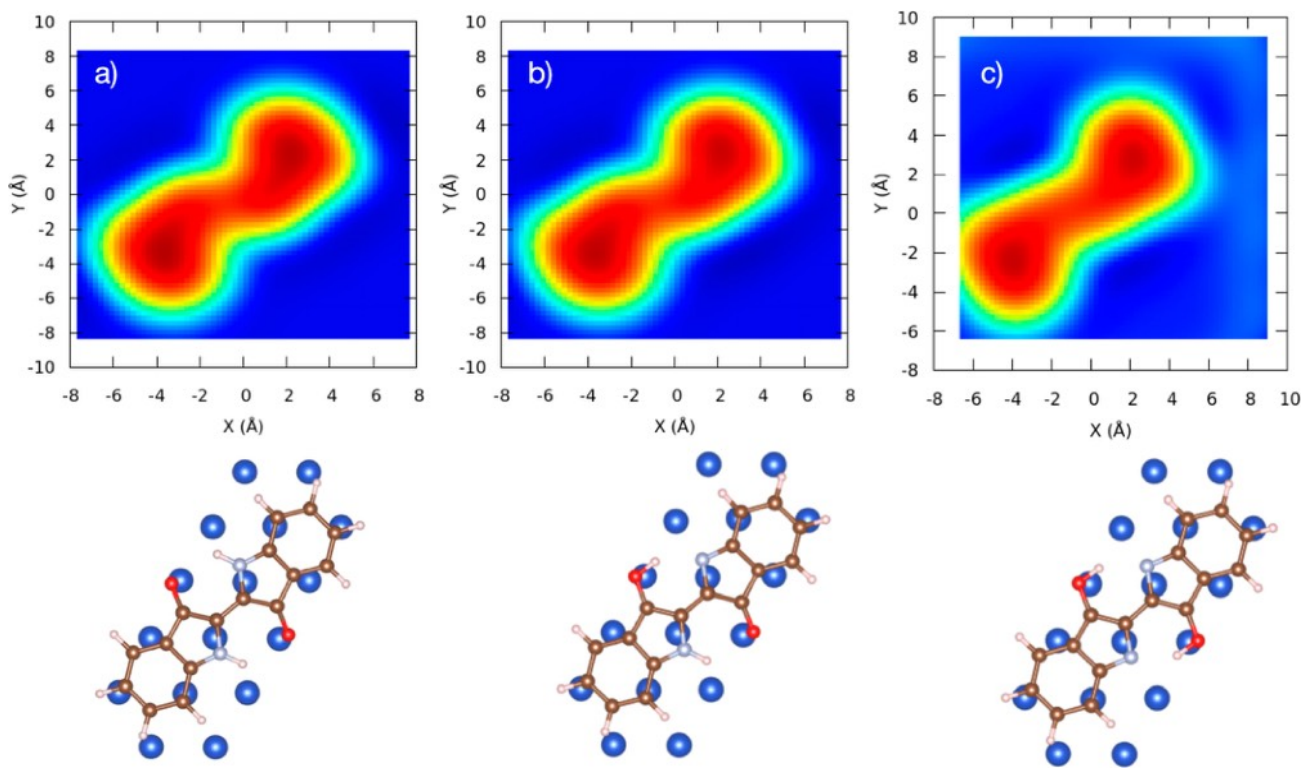

*Figure 6.* Calculated STM images and the corresponding models. (a) Adsorbed molecule, (b) after one proton transfer, and (c) after a double proton transfer (the scanning was displaced 2 Å to the right with respect to (a) and (b) STM-calculated images).

As a second alternative mechanism, we focused on the creation of vacancies in the copper substrate stimulated by the STM pulses, which could be favored by the strong interaction of the oxygen with the copper atoms. Previous experiments of oxygen atoms adsorbed on Cu(111) at room temperature showed that the oxygen atoms upon adsorption release 3-fold copper atoms on the hollow site, producing a preliminary roughness and disorder at the early formation of the $Cu_2O$ layer surface.[102-104] On the other hand, the energy for single surface vacancy formation has been calculated using the embedded-atom method and is about 0.72 eV,[105] whereas DFT-based calculations provided values in the same range: theoretical DFT-D2 calculations with the Grimme method estimated the surface vacancy formation energy on pristine Cu(111) at 0.82 eV,[106] this energy value is higher than 0.69 eV obtained using the DFT-PW91 functional,[107] and it is lower and reasonably similar to other calculated surface vacancy formation energies of 0.92 eV employing the DFT-LDA functional[108] and 1.05 eV using the DFT-PBE functional.[109] The effect of vacancies on the pristine Cu(111) is visualized in the electron charge density map by a depression or hole calculated using the DFT-PBE functional.[109] Notice that once the adatom is on the (111) terrace, it experiences a small diffusion energy: in the literature, one can find calculated hopping barrier values of 28,[105] 34,[110] 41,[111] 44,[112] 50,[113] 53,[114] 59,[115] and 64[115] meV that are comparable to experimental values of 37 ± 5 and 40 ± 1 meV.[116-118] In our model, using the GS geometry as reference, the energy to completely remove the Cu atom below one oxygen atom of the molecule is 2.33 eV, while the energy necessary to remove one Cu atom below a nitrogen atom is 2.73 eV. This energy difference comes from the energy necessary to break the Cu–O bond formed in the first case. The ESQC image of the copper vacancies below the oxygen atom is presented in Figure 7.

Now, the calculated image reproduces qualitatively well the experimental image features: the dark depression at the lateral moiety of the molecule and a slight diminution of the bright lobe belonging to the indoxyl part, where the copper atom of the surface is missing.

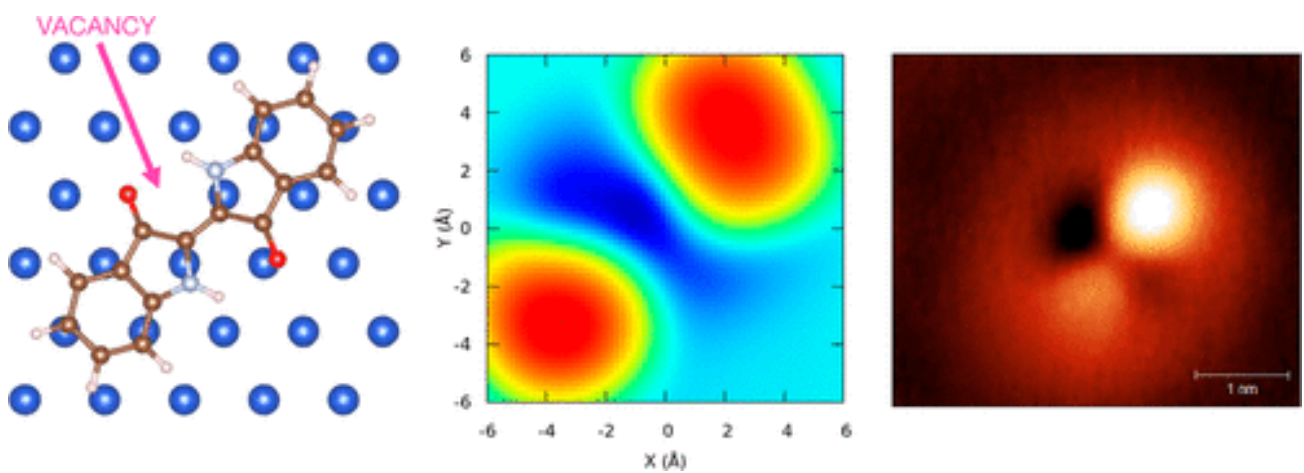

*Figure 7.* Left, the relaxed conformation of the molecule with a single surface vacancy beneath the oxygen atom. Middle, the corresponding calculated STM image. Right, the experimental STM image of a single molecule after a tension pulse (it was rotated for comparison). The pink arrow shows the vacancy and the appearance of a lateral dark spot between two bright spots with different intensities in the calculated image.

In the same way as the single-vacancy calculation, the second Cu vacancy generates a *handle fan-*shaped dark depression on the remaining part of the calculated STM contrast that is expanded at the middle of the molecule, as shown in Figure 8. It is also remarkable to notice that the remaining bright lobe reduces the surrounding bright intensity from the center of the lobe. The experimental STM image after the second voltage pulse is added to the right for comparison in Figure 8. It is observed that both calculated and experimental images match very well.

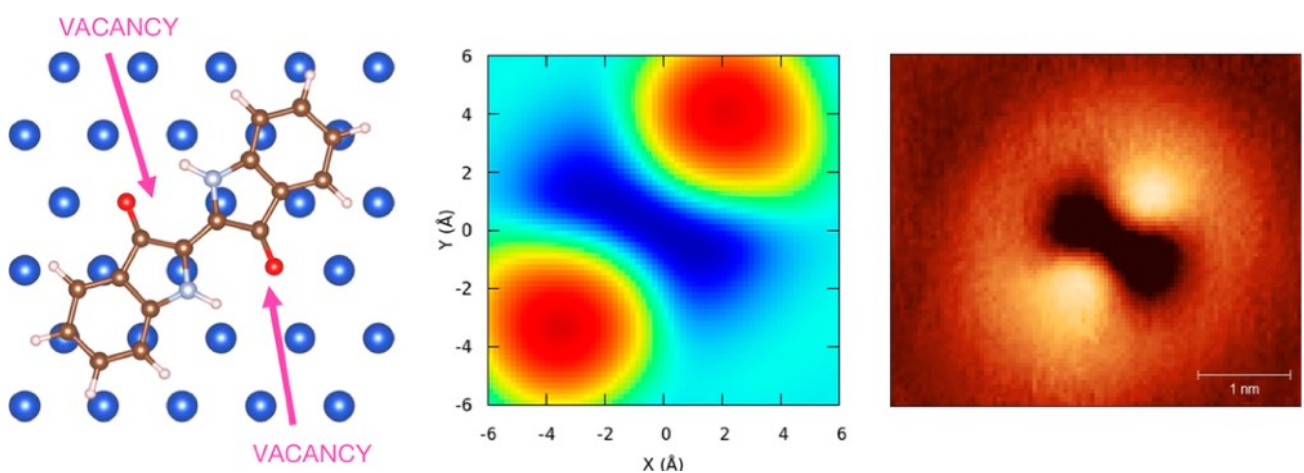

**Figure 8.** Left, the relaxed conformation of the molecule with two surface vacancies beneath the oxygen atoms. Middle, the corresponding calculated STM image. Right, the experimental STM image of a single molecule after a double tension pulse (it was rotated for comparison). Two bright spots are separated by a black channel.

Another investigated mechanism was the creation of a vacancy and the formation of an adatom/molecule complex. Indeed, as mentioned before, the diffusion energy of a single Cu adatom located on the Cu(111) surface is rather small, and consequently the adatom would tend to stay bound to the molecule once extracted from the surface due to a stronger adatom/molecule interaction. DFT calculations assessing this case show that the geometry corresponds to a Cu adatom sitting in a bridge site, bound to two copper surface atoms and interacting with the oxygen atom of the molecule (Figure 9a). The minimal energy geometry of these systems is 1.05 eV higher in energy of the single adsorbed indigo. This new geometry shows a reduction of the Cu–O bond length from 2.14 to 1.90 Å after the total lift of the Cu atom. Bader analysis exhibits an increase of the Cu oxidation, being 0.32 $e^-$ transferred from Cu to the oxygen atom. The interaction of the molecule with the copper substrate generates a new stable configuration that is not possible to find in the gas phase, and it is reached when the Cu adatom of the surface is completely oxidized by the oxygen atom. This state is only possible to be obtained once the Cu atom is pulled out from the surface, and this new state is 0.77 eV higher in energy than the hydrogen transfer. Nevertheless, the oxidation of the copper atom seems to be the most favorable after the introduction of the pulse. These electronic characteristics give the system good energetic stability that in combination with the low-temperature experimental conditions suggest that this configuration could be stable and observed in the STM image. From the geometrical point of view, the N–H, C–N, and C–O bond lengths are 1.02, 1.39, and 1.30 Å, respectively, these values being similar to those of GS, showing that these groups of molecules do not change.

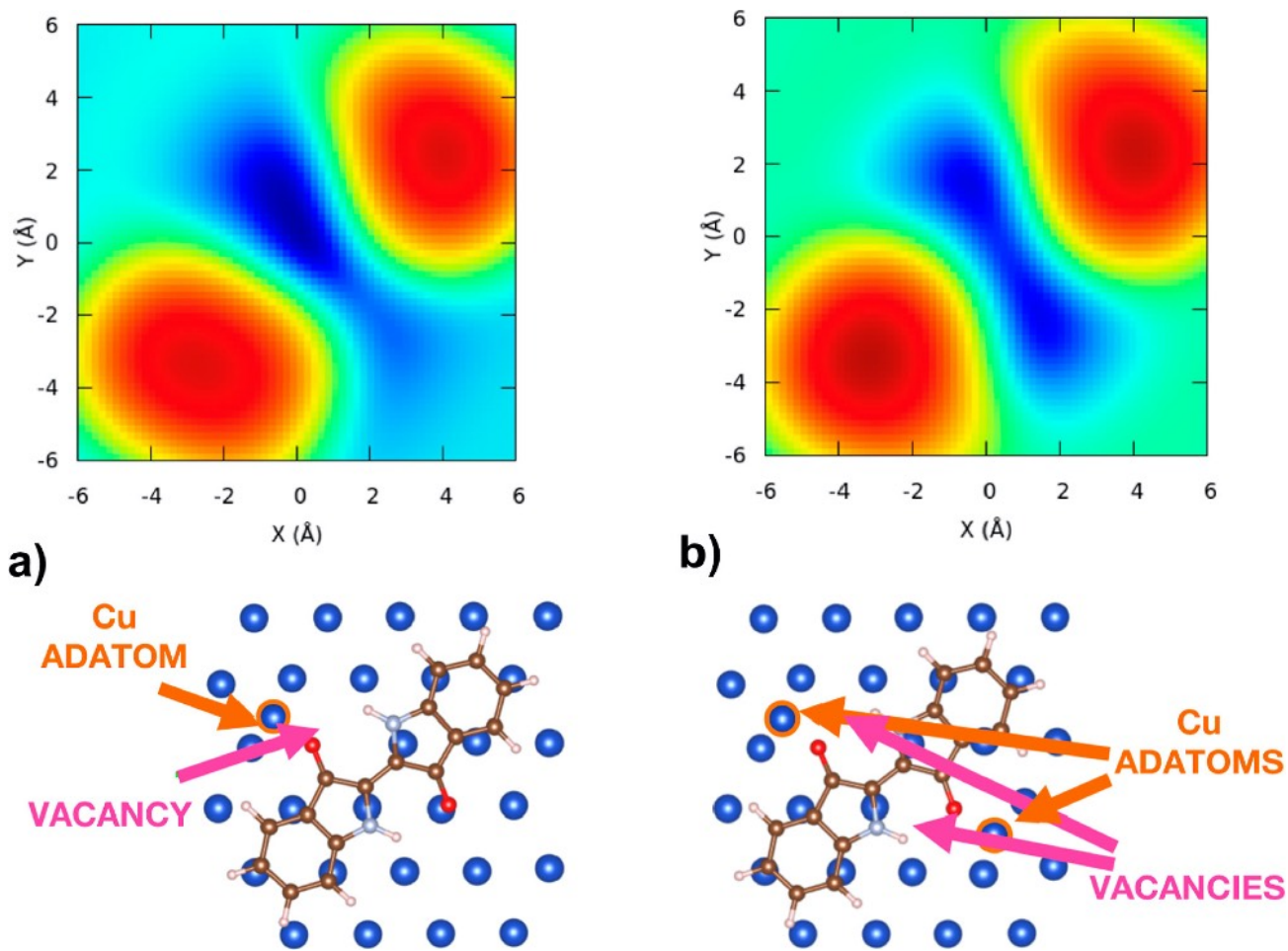

*Figure 9. Left, the calculated STM-ESQC image and the DFT atomic structure of the molecule with one vacancy and the copper adatom extracted from the Cu(111) first layer of the slab. The image shows the depression and almost indistinguishable contrast with respect to the calculated image with a vacancy and without the adatom on the surface (Figure 7). Right, the calculated STM-ESQC image with two vacancies and two adatoms, and the depression is extended at the middle, but has less width than for vacancies alone (Figure 8 ).*

The system with a double lifting of the copper atoms was also analyzed (Figure 9b), and this geometry is less stable than the one lifting, the calculated energy of which is 2.22 eV higher in energy than the ground state. In this case, the two copper atoms present a similar charge transfer toward the oxygen atoms of 0.32 e−, and the Cu–O bond lengths decreased from 2.14 to 1.90 Å as in the one vacancy case.

The calculated STM image of one vacancy and the copper adatom bound to the oxygen atom in Figure 9a also shows the dark depression at the side of the vacancy. The adatom does not seem to show a significant change of the contrast in comparison to the vacancy alone. In the same way, the calculated image of two vacancies at the bottom of the oxygen atoms that are bound to one copper adatom each, as shown in Figure 9b, exhibits the *handle-fan*-shaped dark feature.
The comparison between the two cases, that is to say vacancies and vacancies + adatoms, does not allow us to definitively conclude regarding experimental images. However, DFT calculations would rather show an energetically favorable scenario with the formation of a vacancy generated by an STM voltage excitation with an adatom remaining bound to the molecule.

## Conclusions

We performed STM imaging followed by application of voltage pulses on single indigo molecules adsorbed on Cu(111) at a low temperature, complemented by DFT and ESQC calculations. Experimental STM images showed an irreversible change of contrast with *handle-fan*-shaped dark depression at the middle of the molecule induced by two successive pulses of 2.4 and 3 V in lateral parts of the molecule. In order to explain the experimental results, two main scenarios are suggested. The first one considers proton transfer from the N to O atom induced by the voltage pulses. The TS barrier from the diketo to monoenol forms was calculated by DFT, and it reached about 0.65 eV for a single proton transfer and 0.28 eV for the final geometry. The permanent tautomerization of the single and double proton transfer of N–H···O after two consecutive STM voltage pulses on the molecule is discarded because the comparison between experimental and ESQC calculated images disagrees, and the latter do not show a considerable change of contrast with respect to the indigo adsorbed on the Cu(111) reference. The second scenario is related to the extraction of copper surface atoms by applying successive voltage pulses and creating two vacancies beneath the molecule. Here, experimental and calculated STM images show a remarkable qualitative agreement whether adatoms remain bound to the molecule or not. However, DFT calculations suggest that the adatoms stay close to the molecule and are oxidized by the oxygen atom with 0.32 e$^-$ transfer.

## Author Information

Corresponding Author
- Carlos Javier Villagómez - Instituto de Física, Universidad Nacional Autónoma de México, Apartado Postal 20-364, 01000 Ciudad de México, México; https://orcid.org/0000-0003-0960-0734; Email: cjvillagomez@fisica.unam.mx

Authors
- Fernando Buendía - Instituto de Física, Universidad Nacional Autónoma de México, Apartado Postal 20-364, 01000 Ciudad de México, México

- Lauro Oliver Paz-Borbón - Instituto de Física, Universidad Nacional Autónoma de México, Apartado Postal 20-364, 01000 Ciudad de México, México; https://orcid.org/0000-0002-5086-4558

- Bernardo Fuentes - Instituto de Física, Universidad Nacional Autónoma de México, Apartado Postal 20-364, 01000 Ciudad de México, México

- Tomaso Zambelli - CEMES-CNRS, University of Toulouse 3 Paul Sabatier, 31400 Toulouse, France; Laboratory of Biosensors and Bioelectronics, Institute for Biomedical Engineering, ETH Zürich, 8092 Zürich, Switzerland

- Xavier Bouju - CEMES-CNRS, University of Toulouse 3 Paul Sabatier, 31400 Toulouse, France; https://orcid.org/0000-0001-7827-3496

Notes

The authors declare no competing financial interest.

**Acknowledgments**

C.J.V. acknowledges the project "Estudio de máquinas moleculares y transporte electrónico a la escala atómica" PAPIIT-UNAM IN116320. Part of this work was granted access to the HPC resources of CALMIP Supercomputing Center under the allocation 2021-P0832.